\documentclass[preprint,showpacs,aps,prl,superscriptaddress]{revtex4-1}
\usepackage{bm}
\usepackage{mathrsfs}
\usepackage{amsmath}
\usepackage{amssymb}
\usepackage{graphicx}
\usepackage{amsfonts}
\usepackage{amsthm}
\usepackage{dcolumn}
\usepackage{txfonts}

\begin{document}

\title{Second-order nonlinear optical effects of spin currents}
\author{Jing Wang}
\affiliation{Department of Physics, The Chinese University of Hong Kong, Shatin, N.T., Hong Kong, China}
\affiliation{Department of Physics, Tsinghua University, Beijing 100084, P.R.China}
\author{Bang-Fen Zhu}
\affiliation{Department of Physics, Tsinghua University, Beijing 100084, P.R.China}
\affiliation{Institute of Advanced Study, Tsinghua University, Beijing 100084, P.R.China}
\author{Ren-Bao Liu}
\thanks{To whom correspondence should be addressed. Email: rbliu@phy.cuhk.edu.hk}
\affiliation{Department of Physics, The Chinese University of Hong Kong, Shatin, N.T., Hong Kong, China}

\begin{abstract}
A pure spin current formed by opposite spins moving in opposite directions is a rank-2 axial tensor
which breaks the inversion symmetry. Thus a spin current has a second-order optical susceptibility, with unique
polarization-dependence determined by the symmetry properties of the current. In particular, a longitudinal
spin current, in which the spin polarization directions are parallel or anti-parallel to the moving directions,
being a chiral quantity, leads to a chiral sum-frequency effect. Microscopic calculations based on the
eight-band model of a III-V compound semiconductor confirm the symmetry analysis and show that the susceptibility
is quite measurable under realistic conditions. The second-order nonlinear optical effects
may be used for in-situ and non-destructive detection of spin currents, as a standard spectroscopy tool
in research of spintronics.
\end{abstract}

\maketitle

Spin currents, which carry information via spins in lieu of
charges, play a key role in spintronics~\cite{WolfReview,ZuticReview}.
Pure spin currents also signify the occurrence
of some novel spin-related quantum phenomena such as the spin
Hall effect~\cite{Dyakonov71,Hirsch99PRL,Murakami03Science,Sinova04PRL,Kato04Science,Wunderlich05PRL,Zhao06PRL,Valenzuela06Nature}, the quantum spin Hall effect and topological
insulators~\cite{Kane05QSHE,Bernevig06QSHE,Konig07Science,Hseih09TI3D,Roushan09TI3D}. Spin currents were previously observed via spin
accumulation at stopping edges~\cite{Kato04Science,Wunderlich05PRL,Zhao06PRL,Stevens03PRL} or conversion to electrical
signals~\cite{Valenzuela06Nature,Appelbaum07Nature,Ganichev07PRB,Cui07APL}. Direct and non-destructive measurement of pure
spin currents where and while they flow~\cite{Wang08PRL,Vlaminck08Science} is highly desired,
but is very difficult because a pure spin current bears neither
net charge current nor net magnetization. Noticing that a
longitudinal spin current in which the spins point parallel or
anti-parallel to the current is a chiral quantity, we envisaged
that it can be probed by the chiral sum-frequency optical
spectroscopy which was recently developed to detect molecular
chirality~\cite{ZuticReview2,Wang05CSF,Ji06CSF}. By symmetry analysis in general cases and
microscopic calculations in realistic models, we discovered that
a pure spin current has sizable second-order optical
susceptibility. This finding lays the foundation of direct,
non-destructive measurement of spin currents by standard optical
spectroscopy, facilitating application of spintronics~\cite{WolfReview,ZuticReview} and
research on spin-related quantum phenomena~\cite{Dyakonov71,Hirsch99PRL,Murakami03Science,Sinova04PRL,Kato04Science,Wunderlich05PRL,Zhao06PRL,Valenzuela06Nature,Kane05QSHE,Bernevig06QSHE,Konig07Science,Hseih09TI3D,Roushan09TI3D}.

As a basic principle of nature, a physical object is measurable only when it breaks certain
fundamental symmetries. Indeed, the probe must break the same symmetries as the object does,
since the whole coupled system of an object and a probe has the fundamental symmetries.
For example, in an Amp\`{e}re meter, a ``pure'' charge current, which breaks the time reversal
symmetry, is coupled to a microscopic current inside a magnet. Such symmetry consideration
led to a scheme of detecting a pure spin current by a ``photon spin current'' carried by a
polarized light beam~\cite{Wang08PRL}. A recent experiment~\cite{Vlaminck08Science} showing coupling between a spin
current and a spin wave is a remarkable demonstration of the symmetry principle of measurement.
Though as a direct probe of spin currents, the spin-wave technique~\cite{Vlaminck08Science} still requires special
design and fabrication of magnetic nanostructures and the ``photon spin current'' probe~\cite{Wang08PRL}
is limited by weak interaction since it involves the tiny light momentum, these previous
works paved the way of searching methods of direct and non-destructive measurement of pure
spin currents using symmetry analysis.

\begin{figure}[t]
\centering
\includegraphics[width=\textwidth]{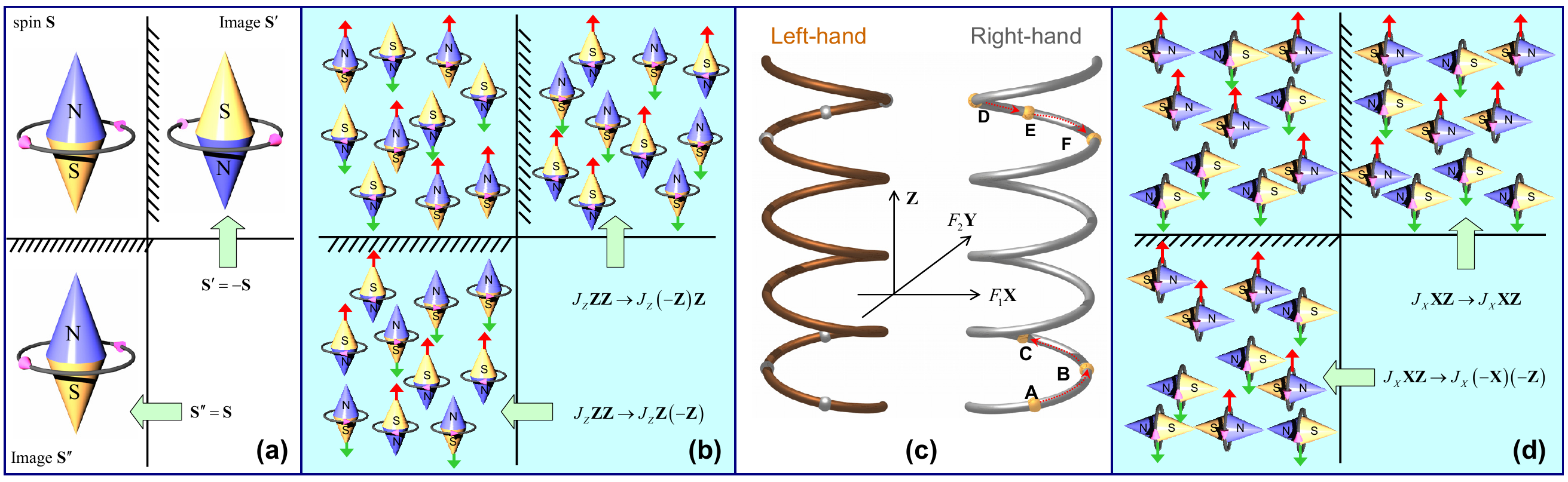}
\caption{(Color online) { Symmetry analysis for sum-frequency effects of spin currents}.
(a) A spin under mirror reflections.
(b) A longitudinal spin current under mirror reflections.
The arrows indicate the moving directions of the spins.
(c) Chiral sum-frequency processes in chiral systems.
(d) A transverse spin current under mirror reflections.} \label{fig1}
\end{figure}

Spin currents have peculiar symmetry properties owing to the characteristics of spins.
Unlike a charge which is a scalar, a spin is a vector pointing to a certain direction.
Physically, a spin is like a tiny magnet resulting from a quantized amount of current circulating
about the spin direction [Fig.~\ref{fig1}~(a)]. Such physical nature makes a spin an unusual vector, namely,
an axial vector. As illustrated in Fig.~\ref{fig1}~(a), a spin reverses inside a parallel mirror and is
unchanged inside a perpendicular mirror, in opposite to a polar vector.
Spin-ups and spin-downs moving in opposite directions with the same velocity make up a pure spin
current without net spin polarization or magnetization. When the spins are parallel or
anti-parallel to the moving direction, the spin current is a longitudinal one.
A longitudinal spin current has a special symmetry property - chirality.
An object, such as a hand or a helix, is chiral if it cannot be made identical to
its mirror image by translation and rotation. The chirality of a longitudinal spin current
is illustrated in Fig.~\ref{fig1}~(b): If a spin's microscopic current circulates its moving direction
left-handedly, the mirror image does right-handedly, and vice versa.

Noticing its chirality, we conceived the idea of measuring a longitudinal spin current
using the chiral sum-frequency optical spectroscopy, which was recently developed as a
standard tool to study molecular chirality~\cite{ZuticReview2,Wang05CSF,Ji06CSF}. In chiral sum-frequency,
two input optical fields ${\mathbf F}_1$ and ${\mathbf F}_2$ (with frequencies $\omega_1$ and $\omega_2$,
respectively) and the induced polarization field ${\mathbf P}$ at frequency $\omega_1+\omega_2$ form a
left- or right-hand system. Fig.~\ref{fig1}~(c) shows how a chiral sum-frequency process occurs in a chiral system.
Considering a right-hand helix, a charge at position $A$ will be driven to point $B$ by an electric field ${\mathbf F}_1$
which is along the $X$-axis, and then to point $C$ by ${\mathbf F}_2$ which is along the $Y$-axis.
The confinement of the helix leads to a net displacement along the $Z$-axis. Thus the two input fields
and the induced polarization $({\mathbf F}_1, {\mathbf F}_2, {\mathbf P})$ form a right-hand system.
If the order of input fields is reversed (${\mathbf F}_2$ applies before ${\mathbf F}_1$),
the charge would follow a trajectory like $D\rightarrow E\rightarrow F$, resulting in a polarization along the $-Z$-axis,
and $({\mathbf F}_2, {\mathbf F}_1, {\mathbf P})$ still form a right-hand system.
Similarly, the sum-frequency in a left-hand helix is a left-hand chiral process.
A sum-frequency process is characterized by a second-order susceptibility $\chi^{(2)}$
via ${\mathbf P}\left(\omega_1+\omega_2\right)=\chi^{(2)}:{\mathbf F}_1\left(\omega_1\right){\mathbf F}_2\left(\omega_2\right)$.
In the above example of helix, the susceptibility may be written as a form of three dyadic vectors,
$\chi^{(2)}=A\left(\mathbf{ ZYX}-\mathbf{ ZXY}\right)$, i.e., a rank-3 tensor.
Thus the chiral sum-frequency susceptibility provides a measurement
of the chirality of a physical object. If otherwise measured in linear optics, the effect of the molecular
chirality relies on the small magnetic moment of the molecules, and in turn on the small photon momentum
of the probe light, similar to the case of linear optical effects of spin currents~\cite{Wang08PRL}.

For a systematic symmetry analysis, we consider a spin current with both longitudinal and transverse components.
We define the $Z$-axis as the current direction and the $X$-axis as the spin direction of the transverse component.
The spin current can be written as a rank-2 tensor ${\mathbb J}=J_X\mathbf{ XZ}+J_Z\mathbf{ ZZ}$,
in a form of dyadic vectors, in which the left/right vector is the spin/current direction and $J_{Z/X}$
is the longitudinal/transverse amplitude. Above all, the spin current breaks the inversion symmetry,
satisfying the symmetry properties required by a second-order optical process~\cite{Shen_NO}.

In general, the sum-frequency susceptibility tensor has 27 independent terms,
$\chi^{(2)}=\chi_{XXX}\mathbf{ XXX}+\chi_{XXY}\mathbf{ XXY}+\cdots+\chi_{ZZZ}\mathbf{ ZZZ}$,
but the symmetry properties of a spin current will set many terms to be zero or non-independent~\cite{Shen_NO}.
For a longitudinal spin current, only the chiral terms are non-zero. In a non-chiral term,
at least one of the three directions $\mathbf{ X}$, $\mathbf{ Y}$ and $\mathbf{ Z}$ appears even times
(twice or zero times). Consider $\chi_{XXY}\mathbf{ XXY}$ for example.
Under reflection by the $Y$-$Z$ plane, the longitudinal spin current is reversed, but $\chi_{XXY}\mathbf{ XXY}$
is unchanged, so this term must be zero. Similar arguments apply to other non-chiral terms.
Also, the susceptibility must be anti-symmetric under reflection by any plane parallel to the $Z$-axis.
With these constraints, the sum-frequency susceptibility of a longitudinal spin current can be written as
\begin{align}
\chi^{(2)}_{J_Z}=J_Z\Big[& \alpha_1\left(\mathbf{ XYZ}-\mathbf{ YXZ}\right)
\nonumber \\
+ & \alpha_2\left(\mathbf{ YZX}-\mathbf{ XZY}\right)
+\alpha_3\left(\mathbf{ ZXY}-\mathbf{ ZYX}\right)\Big],
\label{Eq_chiJz}
\end{align}
with only three independent parameters. As for a transverse spin current $J_X\mathbf{XZ}$,
it changes its sign under reflection by the $X$-$Z$ plane but is invariant under reflection by the
$X$-$Y$ or $Y$-$Z$ plane [see Fig.~\ref{fig1}~(d)], each non-zero term in the susceptibility must contain odd times of
${\mathbf Y}$ and even times of ${\mathbf Z}$ or ${\mathbf X}$, so
\begin{align}
\chi^{(2)}_{J_X}=J_X\Big(&x_1\mathbf{XXY}+x_2\mathbf{XYX}+x_3\mathbf{YXX}
\nonumber \\
+ & z_1\mathbf{ZZY}
+z_2\mathbf{ZYZ}+z_3\mathbf{YZZ}+y\mathbf{YYY}\Big),
\label{Eq_chiJx}
\end{align}
with seven independent parameters.
The unique polarization-dependence of the second-order susceptibility of a spin current
can be used to distinguish its transverse and longitudinal components, and also to
single out the spin-current signature from the effects of the material background
or a charge current~\cite{Khurgin95}.

\begin{figure}[t]
\centering
\includegraphics[width=\textwidth]{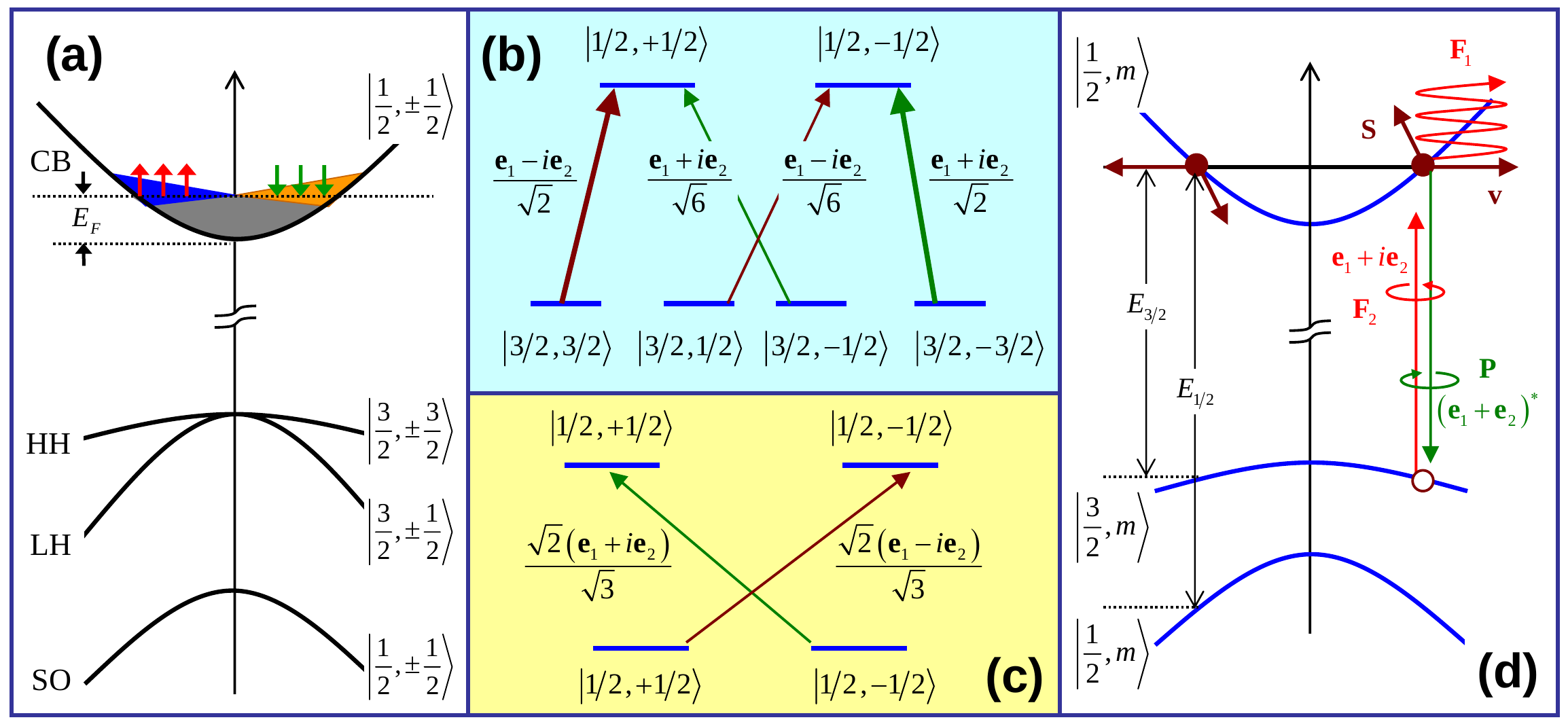}
\caption{(Color online) {Models for microscopic calculation of the sum-frequency susceptibility.}
(a) The full eight-band model and the electron spin distribution for a pure spin current in a semiconductor.
(b) and (c) Selection rules and relative dipole moments from the spin-3/2 and spin-1/2 valence bands to
the conduction band, respectively. (d) A simplified model with the HH-LH splitting neglected.
The spin states and selection rules for inter-band transitions are independent of the momentum.
The transition energies to the Fermi surface from different valence bands are indicated.}
\label{fig2}
\end{figure}

\begin{figure}[t]
\begin{center}
\includegraphics[width=\textwidth]{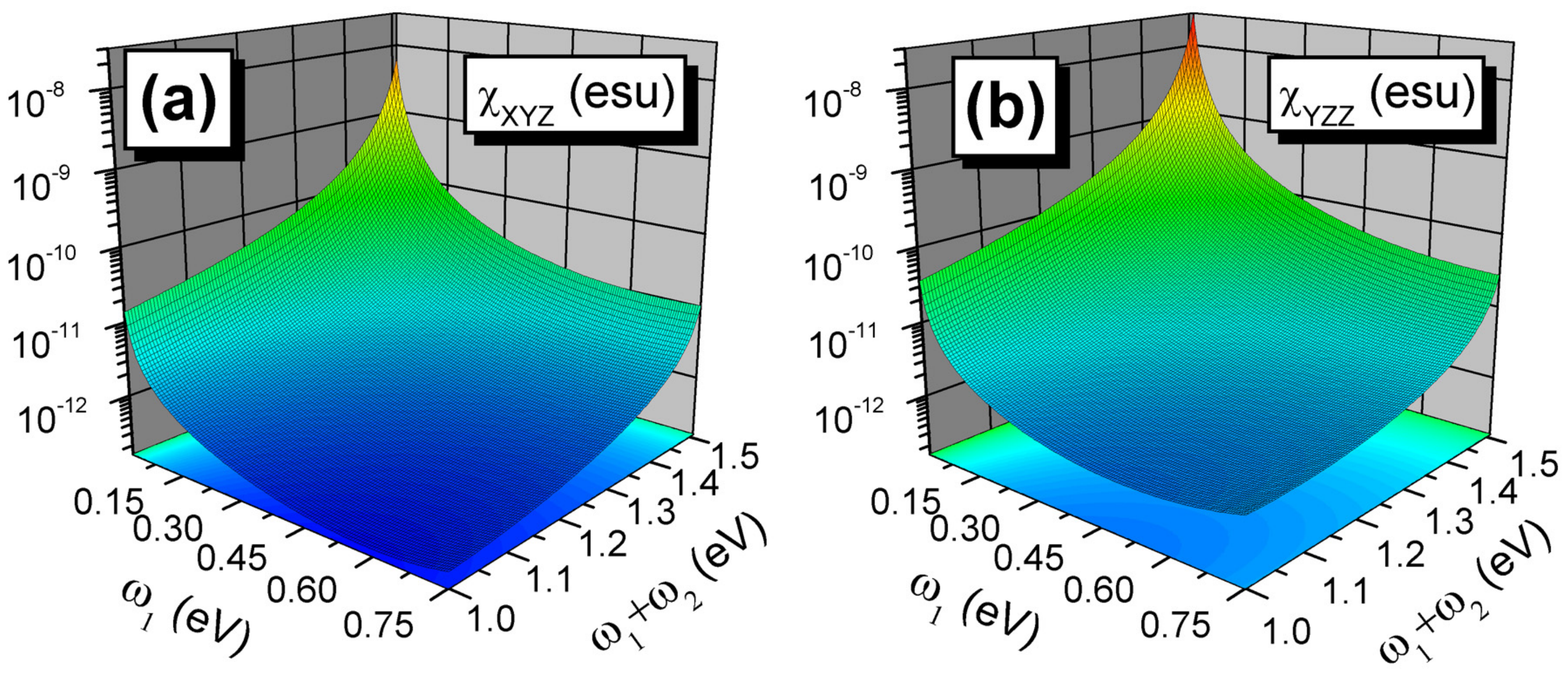}
\end{center}
\caption{(Color online) { Representative results of the sum-frequency susceptibility}.
(a) $\chi_{XYZ}$ due to a longitudinal spin current, and
(b) $\chi_{YZZ}$ due to a transverse spin current, as functions of the optical frequencies.
Parameters are chosen similar to those in Ref.~\cite{Kato04Science}:
The band gap is 1519~meV, the HH-SO splitting is 341~meV, the doping concentration is $3\times 10^{16}$~cm$^{-3}$,
the effective mass (in units of free electron mass) of the HH, LH, SO, and conduction bands is in turn 0.45, 0.082, 0.15,
and 0.067, the dipole $d_{\rm cv}=6.7$~e\AA, the dielectric constant $\varepsilon_r=10.6$,
and the spin current $J_X=J_Z=20$~nA/$\mu$m$^2$.}
\label{fig3}
\end{figure}

To determine the independent parameters of the susceptibility in Eqns.~(\ref{Eq_chiJz}) and (\ref{Eq_chiJx}),
we performed microscopic calculation for a pure spin current in a bulk GaAs,
using the standard perturbation theory~\cite{Shen_NO,Sipe_NO} with an eight-band model~\cite{YuSemi}.
We assumed that the pure spin current result from a non-equilibrium distribution of electrons
in the conduction band, with a small portion of electrons near the Fermi surface having opposite
spin polarizations for opposite velocities [Fig.~\ref{fig2}~(a)] , under conditions similar to those in Ref.~\cite{Kato04Science}.
The optical interaction includes the inter-band transitions between the valence bands and the
conduction band and the intra-band acceleration of electrons and
holes. To avoid real absorption of light, the light frequencies were
chosen such that the sum frequency is below the band gap. For the
sake of simplicity, we neglected the anisotropy of the valence
bands. We also adopted the free-particle approximation, which is
justified since the Coulomb interaction is largely screened in
the n-doped material. These approximations, according to the
symmetry analysis, would only quantitatively modify the results.
The spin splitting due to the bulk inversion asymmetry of the
material (the Dresselhaus effect) is as small as 0.01~meV for the
doping level considered ($3\times 10^{16}$~cm$^{-3}$), and
therefore was neglected in the calculation. The bulk inversion
asymmetry would cause a background second-order susceptibility,
which is indeed strong but can be well separated from the spin-current effect
by AC modulation of the current and phase-locking detection.
Two representative results of the calculated susceptibility spectra are shown in Figs.~\ref{fig3}~(a) and (b).
The other terms of the susceptibility tensor (not shown) have similar frequency-dependence and
comparable amplitudes.
As a specific example, a transverse spin current 20~nA/$\mu$m$^{-2}$ has a
susceptibility $\chi_{YZZ}\approx 0.40\times 10^{-9}$~esu (or $0.17\times 10^{-12}$~m/V in SI units)
for input frequencies $\omega_1=100$~meV and $\omega_2=1,400$~meV, or $17.\times 10^{-12}$~esu
for $\omega_1=\omega_2=750$~meV (corresponding to the second harmonics generation).

To better understand the microscopic mechanism of the sum-frequency effect of a spin current,
we simplify the model by neglecting the splitting between the heavy hole (HH) band and the light hole
(LH) band. Under this approximation, the HH and LH bands form a spin-3/2 band with 4-fold degeneracy.
The split-off (SO) band and the conduction band have spin-1/2. In this simplified model,
the spin states and the selection rules for inter-band transitions are separated from the momentum
[Figs.~\ref{fig2}~(b) and (c)].

Let us first consider a single electron with momentum ${\mathbf k}$ and spin polarization ${\mathbf s}_{\mathbf k}$
[Fig.~\ref{fig2}~(d)]. We set up a coordinate system $({\mathbf e}_1, {\mathbf e}_2, {\mathbf e}_3)$ so that
${\mathbf s}_{\mathbf k}={\mathbf e}_3\left(f_+ -f_-\right)/2$ with $f_{+/-}$ denoting the population at the
spin-up/down state. The angular momentum conservation requires that a light with circular polarization
${\mathbf e}_1\pm i{\mathbf e}_2$ couples only to the transitions $|j,m\rangle\leftrightarrow|1/2,m\pm 1\rangle$,
where $j=3/2$ or $1/2$ is the spin of a valence band and $m=-j$, $j+1$, $\ldots$, or $j$ is
the component along the ${\mathbf e}_3$-axis. The relative dipole moments of the relevant inter-band
transitions are indicated in Figs.~\ref{fig2}~(b) and (c). To simplify the discussion, we set the input frequency $\omega_2$ to be near resonant with the band gap and much greater than $\omega_1$, so that the inter-band transitions and the intra-band driving are mostly caused by ${\mathbf F}_2\exp\left(-i\omega_2t_2\right)$ and
${\mathbf F}_1\exp\left(-i\omega_1t_1\right)$, respectively.

The probability amplitude of a certain inter-band transition is determined by the inner product of the dipole moment and the optical field. For example, the transition
$|3/2,-3/2\rangle\rightarrow |1/2,1/2\rangle$ generated between $t_2$ and $t_2+dt_2$
has a probability amplitude $dG_2=i(1-f_-)\left(d_{\rm cv}^*/\sqrt{2}\right)\left({\mathbf e}_1+i{\mathbf e}_2\right)
\cdot{\mathbf F}_2\exp(-i\omega_2t_2)dt_2$,
where $d_{\rm cv}$ is the inter-band dipole, and the factor $(1-f_-)$ accounts for the Pauli blocking.
After the excitation, the probability amplitude oscillates in time with frequency $E_{3/2}({\mathbf k})$,
leading to the optical polarization $\left({\mathbf e}_1+i{\mathbf e}_2\right)\left(d_{\rm cv}/\sqrt{2}\right)
e^{-iE_{3/2}({\mathbf k})(t-t_2)}dG_2$
at time $t$, where $E_{3/2}(\mathbf{k})=k^2/(2m_e)+k^2/(2m_{3/2})$ is the transition energy of a pair of
electron an hole with mass $m_e$ and $m_{3/2}$, respectively. The radiation has the same circular polarization
as the input because of the angular momentum conservation. Summation over all possible transitions and
integration over time give the linear optical response to the field $\mathbf{F}_2$ as
\begin{align}
{\mathbf P}^{(1)}(t)&=\frac{i}{3}\left|d_{\rm cv}\right|^2\int_{-\infty}^te^{-iE_{3/2}(\mathbf{k})(t-t_2)}
\nonumber \\
& \times
\sum_{\pm}\left(1-f_{\pm}\right)\left(\mathbf{e}_1\mp i\mathbf{e}_2\right)\left(\mathbf{e}_1\mp i\mathbf{e}_2\right)^*
\cdot\mathbf{F}_2e^{-i\omega_2t_2}dt_2.
\label{Eq_P1}
\end{align}
Thus  $\mathbf{P}^{(1)}\propto s_{\mathbf k}\left(\mathbf{e}_1\mathbf{e}_2-\mathbf{e}_2\mathbf{e}_1\right)\cdot\mathbf{F}_2
=\mathbf{F}_2\times\mathbf{s}_{\mathbf k}$, which has a transparent physical meaning: The linear polarization
of the output field is related to that of the input one by a rotation about the spin,
essentially a Faraday rotation due to the spin acting as a magnet.
When the effect of the intra-band driving by $\mathbf{F}_1$ is included, the momentum $k$ should be
replaced with the accelerated one $\tilde{\mathbf k}_{\tau}
\equiv \mathbf{k}-e\mathbf{F}_1\int_{-\infty}^{\tau} \exp(-i\omega_1 t_1)dt_1$ at time $\tau$,
and the phase $E_{3/2}(\mathbf{k})(t-t_2)$ accumulated from the creation time $t_2$ to the recombination time $t$
should be replaced with $\int_{t_2}^tE_{3/2}\left(\tilde{\mathbf k}_{\tau}\right)d\tau$.
By expansion to the linear order of $\mathbf{F}_1$, we have
$\tilde{k}_{\tau}^2\approx k^2-2e{\mathbf k}\cdot\mathbf{F}_1\int_{-\infty}^{\tau}\exp(-i\omega_1t_1)dt_1$,
so the second-order optical response can be written as
$\mathbf{P}\propto\mathbf{F}_2\times\mathbf{s}_{\mathbf k}e{\mathbf v}_{\mathbf k}\cdot\mathbf{F}_1$,
where $\mathbf{v}_{\mathbf k}\equiv{\mathbf k}/m_e$ is the velocity of the electron with momentum $\mathbf{k}$.
The physical meaning of  $e\mathbf{v}_{\mathbf k}\cdot\mathbf{F}_1$ is obviously the power
done by the field to the electron. $e\mathbf{s}_{\mathbf k}\mathbf{v}_{\mathbf k}$ is just the spin current
tensor contributed by the electron.

For a distribution of electrons, the summation over the momentum space gives the sum-frequency response as
$\mathbf{P}=\zeta\mathbf{F}_2\times\left({\mathbb J}\cdot{\mathbf F}_1\right)$, with
\begin{align}
\zeta=& \left(\frac{\varepsilon_r+2}{3}\right)^3\frac{\left(2/3\right)\left|d_{\rm cv}\right|^2\left(1+m_e/m_{3/2}\right)}
{\left(\omega_1+\omega_2-E_{3/2}\right)\left(\omega_2-E_{3/2}\right)\omega_1} \nonumber \\
& -
\Bigg(E_{3/2},m_{3/2}\rightarrow E_{1/2},m_{1/2}\Bigg),
\label{Eq_zeta}
\end{align}
derived by Fourier transformation of Eq.~(\ref{Eq_P1}) including the intra-band driving and the contribution of the SO band,
where the factor containing the material dielectric constant $\varepsilon_r$ takes into account the difference
between the macroscopic external field and the microscopic local field~\cite{Bloembergen65NO}, $m_j$ denotes the mass of the spin-$j$
hole band, and $E_j$ is the transition energy from the spin-$j$ band to the Fermi surface [see Fig.~\ref{fig2}~(d)].
The constants in Eqns.~(\ref{Eq_chiJz}) and (\ref{Eq_chiJx}) are such that $\alpha_1=-z_2=z_3=\zeta$ and others$=0$.
With the HH-LH splitting neglected, the sum-frequency susceptibility has a compact form with only
one independent parameter. This feature is due to the separation of the spin and motion degrees of
freedom of the electrons and holes. When the HH-LH splitting is non-zero, the spin quantization direction
and therefore the optical selection rules depend on the momentum and vary with acceleration of the particles.
This leads to the general form of susceptibility in Eqns.~(\ref{Eq_chiJz}) and (\ref{Eq_chiJx}), with the extra terms proportional to
the HH-LH splitting.

In summary, with systematic symmetry analysis in general cases and microscopic calculation under
realistic conditions, we have shown that a pure spin current has a sizable sum-frequency susceptibility.
In particular, a longitudinal spin current has a chiral sum-frequency effect.
The current results can be straightforwardly extended to other second-order
optical spectroscopy such as difference-frequency and three-wave mixing~\cite{Shen_NO}.
Thus the standard nonlinear optical spectroscopy makes up a toolbox for research of spintronics.
With universality of the method guaranteed by the symmetry principle and without requirements of
resonance conditions or special structure design and fabrication, the nonlinear optical spectroscopy
can be applied to study a wide range of spin-related quantum phenomena such as the quantum spin Hall
effect and topological insulators~\cite{Kane05QSHE,Bernevig06QSHE,Konig07Science,Hseih09TI3D,Roushan09TI3D}.
A wealth of physics connecting spins and photons and
technologies synthesizing spintronics and photonics are still to be explored.

This work was supported by Hong Kong RGC HKU 10/CRF/08, Hong Kong GRF CUHK 402207,
the NSFC Grant Nos. 10774086, 10574076 and the Basic Research Program of China Grant No. 2006CB921500.

\end{document}